\documentclass[10pt,journal,compsoc]{IEEEtran}
\ifCLASSOPTIONcompsoc
  \usepackage[nocompress]{cite}
\else
  \usepackage{cite}
\fi
\ifCLASSINFOpdf
   \usepackage[pdftex]{graphicx}
   \graphicspath{{../pdf/}{../jpeg/}}
   \DeclareGraphicsExtensions{.pdf,.jpeg,.png}
\else
   \usepackage[dvips]{graphicx}
   \graphicspath{{../eps/}}
   \DeclareGraphicsExtensions{.eps}
\fi
\begin{document}
%
\title{Facial movement synergies and Action Unit detection from distal wearable Electromyography and Computer Vision}
%
%
%
%

\author{Monica~Perusqu\'ia-Hern\'andez,~\IEEEmembership{Member,~IEEE, }
        Felix Dollack,~\IEEEmembership{Member,~IEEE, }
        Chun Kwang Tan,
        Shushi Namba,
        Saho Ayabe-Kanamura,
        and~Kenji~Suzuki,~\IEEEmembership{Member,~IEEE}
\IEEEcompsocitemizethanks{\IEEEcompsocthanksitem M.P-H. and F.D. are with NTT Communication Science Laboratories\protect\\
E-mail: monica.perusquia.mp@hco.ntt.co.jp; perusquia@ieee.org %
\IEEEcompsocthanksitem C.K.T., S.A-K., and K.S. are with the University of Tsukuba
\IEEEcompsocthanksitem S.N. is with the University of Hiroshima}
\thanks{Manuscript prepared January, 2020.}}

%
%

\markboth{Journal of \LaTeX\ Class Files,~Vol.~14, No.~8, August~2020}%
{Shell \MakeLowercase{\textit{et al.}}: Bare Demo of IEEEtran.cls for Computer Society Journals}
%



\IEEEtitleabstractindextext{%
\begin{abstract}
Distal facial Electromyography (EMG) can be used to detect smiles and frowns with reasonable accuracy. It capitalizes on volume conduction to detect relevant muscle activity, even when the electrodes are not placed directly on the source muscle. The main advantage of this method is to prevent occlusion and obstruction of the facial expression production, whilst allowing EMG measurements. However, measuring EMG distally entails that the exact source of the facial movement is unknown. We propose a novel method to estimate specific Facial Action Units (AUs) from distal facial EMG and Computer Vision (CV). This method is based on Independent Component Analysis (ICA), Non-Negative Matrix Factorization (NNMF), and sorting of the resulting components to determine which is the most likely to correspond to each CV-labeled action unit (AU). Performance on the detection of AU06 (Orbicularis Oculi) and AU12 (Zygomaticus Major) was estimated by calculating the agreement with Human Coders. The results of our proposed algorithm showed an accuracy of 81\% and a Cohen's Kappa of 0.49 for AU6; and accuracy of 82\% and a Cohen's Kappa of 0.53 for AU12. This demonstrates the potential of distal EMG to detect individual facial movements. Using this multimodal method, several AU synergies were identified. We quantified the co-occurrence and timing of AU6 and AU12 in posed and spontaneous smiles using the human-coded labels, and for comparison, using the continuous CV-labels. The co-occurrence analysis was also performed on the EMG-based labels to uncover the relationship between muscle synergies and the kinematics of visible facial movement.
\end{abstract}

}

\maketitle

\IEEEdisplaynontitleabstractindextext

\begin{figure}[!t]
\centering
\includegraphics[width=2.5in]{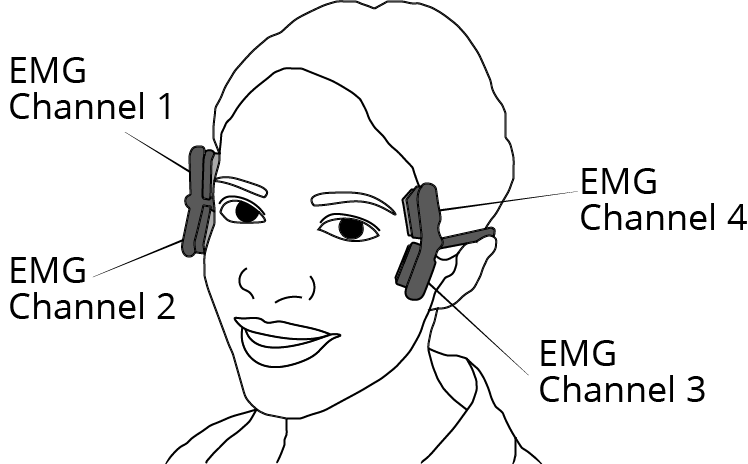}
\caption{Wearable used to measure distal EMG from four channels placed on the sides of the face, on both temples of the head. This configuration enables facial expression identification without obstructing the face. However, this makes identifying which muscle produced the measured activity challenging.}
\label{fig:wearable}
\end{figure}

\begin{figure}[!t]
\centering
\includegraphics[width=2.5in]{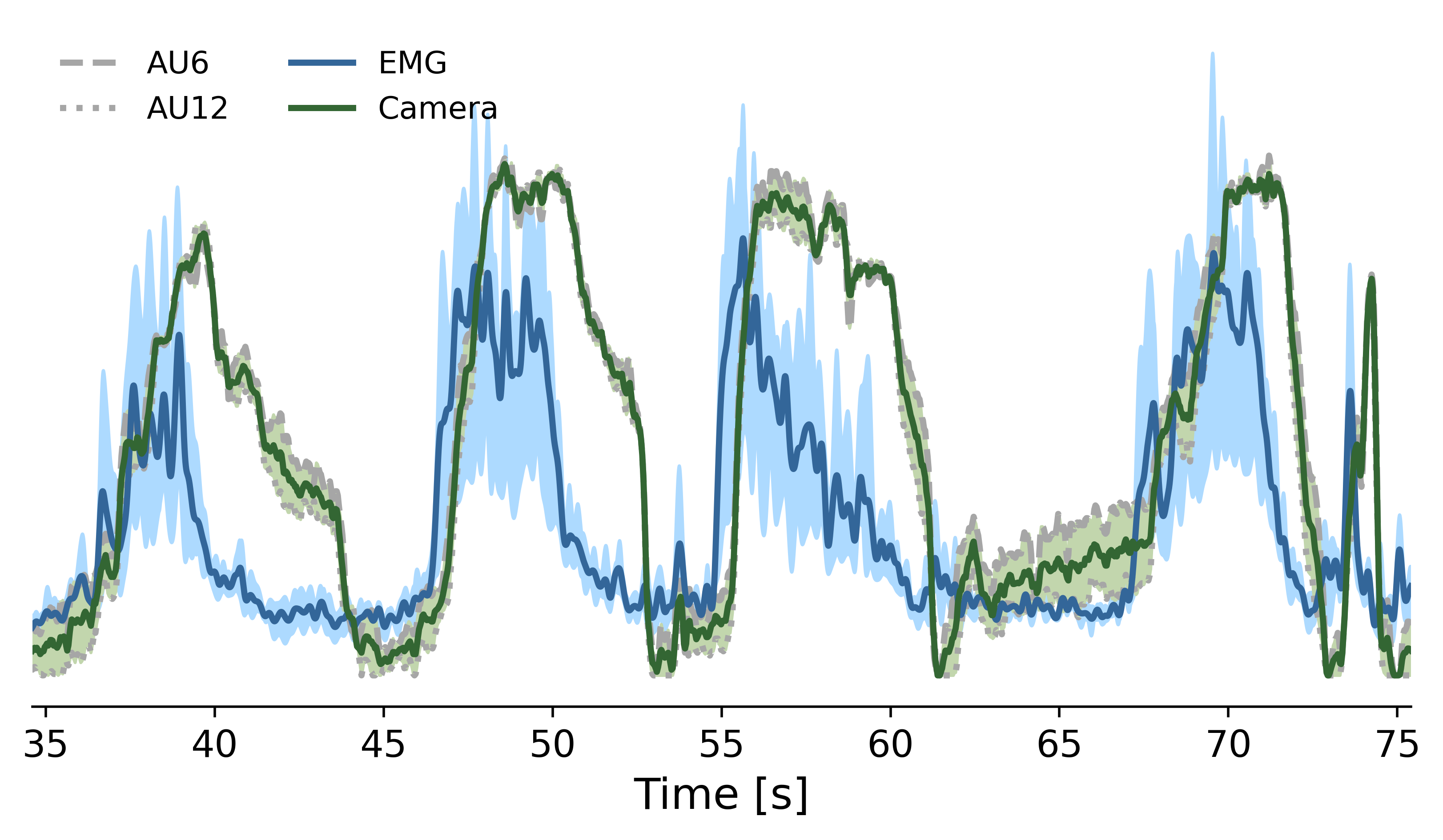}
\caption{Muscle activations measured with distal EMG precede camera-detected AU movements. The four channels of raw distal EMG activate on average 374~ms before the detected CV-based AU labels. The plot shows data from one participant posing smiles. The activation patterns of EMG and CV-based AUs are similar to each other, with EMG activity leading. The blue line shows the mean of the four EMG channels plus Standard Deviation (SD). AU6 and AU12 often co-occur, as shown by the CV-based output. This makes identifying which muscle produced the measured activity challenging, as EMG measures a mix of muscle activity throughout the face.}
\label{fig:processing}
\end{figure}

%
\IEEEpeerreviewmaketitle

\IEEEraisesectionheading{\section{Introduction}\label{sec:introduction}}

\IEEEPARstart{F}{acial} expressions often co-occur with affective experiences. Smiles are among the most ubiquitous facial expressions. They are characterized by the corner of the lips moving upwards, as a resulting action of the activity of the \textit{Zygomaticus Major} muscle (ZM). Although smiles have been deemed the prototypical expression of happiness, not all people who smile are happy. The so-called Duchenne marker, or movement from the \textit{Orbicularis Oculi} muscle (OO), often co-occurs with the \textit{Zygomaticus Major} activity. Whilst it has been claimed that the Duchenne marker is a signal of smile spontaneity~\cite{Ekman1988,Ekman1988a,Ekman1999,Guo2018}, other studies have found this marker in posed smiles as well~\cite{Schmidt2009,Krumhuber2009,Namba2016}. Others suggested that it might also signal smile intensity instead of smile authenticity~\cite{Messinger2002,Krumhuber2009,Girard2019}. 

The Facial Action Coding System (FACS)~\cite{Ekman1972} is a standardized method to label facial movements. It involves identifying Action Descriptors for movements involving multiple muscles from their onset to offset. The advantage of using the FACS is that no subjective inferences about the underlying emotion are made during the facial movement identification. From these AU configurations, inferences can be made by experts in the frame of different theories of emotion. In the FACS, the lip corner pulling upwards is labeled as Action Unit 12 (AU12), and the movement around the eyes in the form of a cheek raiser is labeled as AU6. AU6 is also the AU associated with the Duchenne Marker. AUs can be measured by visual inspection using video recordings, either by a human coder~\cite{Ekman2002} or by using Computer Vision (CV) algorithms~\cite{Baltrusaitis2018}. Additionally, the underlying muscle activity can be measured with Electromyography (EMG)~\cite{VanBoxtel2010,Schmidt2001}. The standard method is to place the EMG electrodes directly on top of the relevant muscle to increase Signal-to-Noise Ratio (SNR). More recently, several studies have proved the feasibility of measuring facial expressions with distal EMG~\cite{Hamedi2013,Gruebler2014}. Distal EMG refers to measuring muscle activity from a body location that is distant from the relevant muscle. Distal EMG measurements are possible through volume conduction whereby the electrical activity generated by each muscle spreads to adjacent areas~\cite{VanBoxtel2010}. By measuring EMG distally, the unnatural obstruction that the electrodes pose to the production of facial expressions is reduced. Despite this advantage, distal measurements make it difficult to know the exact location of the EMG activity source. Hence, current technology has been only used to identify grouped muscle activity such as smiles or frowns. Detecting such facial expressions from EMG has its own merit, such as high temporal resolution and robustness against occlusion. However, to compare the knowledge drawn using this technology to the large body of facial expression research that uses AUs as the basis of analysis, we need to identify muscle movement activity at the AU level. 

Units of movement are often grouped to form full facial expressions. In movement science, muscle synergies are defined as joint movements produced by muscle groups~\cite{Tresch2009}. Analogously, we define \textit{facial movement synergies} as groups of muscles, or AUs, moving together. We hypothesize that muscle synergies might not correspond 1:1 to visible observable synergies, due to the differences between muscle activity and the kinematics of the movement. Additionally, by using synergy analysis, we should be able to observe different synergies involving AU6 and AU12 in the context of posed and spontaneous smiles. This might provide more evidence in favor or against AU6 being a marker of spontaneity. If AU6 and AU12 move together in spontaneous smiles, but not in posed smiles, the Duchenne marker might be related to smile genuineness. On the other hand, if the Duchenne marker is not a signature of spontaneity, we should observe similar synergies in both posed and spontaneous smiles. We conducted this co-occurrence analysis using human-labeled AUs, CV-labeled AUs, and EMG-labeled AUs. The objective was to find common factors in how a smile is produced, and the relationship between visible movement and EMG synergies.

For the purpose of this analysis, we propose a sensing-source-synergy framework. Blind Source Separation (BSS) methods, such as Independent Component Analysis (ICA) or pre-trained OpenFace models, could be used to go from sensing to sources. Furthermore, analyses such as Non-Negative Matrix Factorization (NNMF) can be used to identify synergies from fine-grained movement units. We explore several algorithms within this space to shed light on the spatial and temporal elements that form posed and spontaneous smiles.

Moreover, it is possible to create an AU identification system that works in recording sessions where high movement or high facial occlusion are expected by combining CV- and EMG- based methods. In those cases, CV alone would struggle to continuously identify certain AUs. On the other hand, wearable distal EMG can deal with occlusion and movement, but it cannot disentangle AUs so easily. With our proposed method, CV can be used in a calibration period as a reference for automatic AU identification in EMG. Thus, we would be able to create a more robust system that enables fine-grained analysis of facial expression synergies and activation. A use case scenario would be when assessing children's behavior, or when developing applications for Virtual Reality. It would be possible to have a short calibration session with both camera and EMG recordings. Using this multimodal information, the EMG could be tagged and used to identify AUs without requiring constant camera surveillance while the children play and run around, or when the face is covered by a headset.

The main contributions of this work are:
\begin{enumerate}
    \item \textbf{A framework} to analyze facial activity by estimating sensed signal sources and synergies.
    \item \textbf{A method to identify individual muscle activity sources} linked to the AUs 6 and 12 during smile production from a multimodal system. This system uses both CV and EMG during calibration, and EMG only for high-movement, high-occlusion situations. We use CV as reference method to identify different EMG components. This method is based on Independent Component Analysis (ICA) for Blind Source Separation (BSS) of the EMG readings to estimate the source muscle. Additionally, we use Non-Negative Matrix Factorization (NNMF) to identify AU synergies. 
    \item \textbf{An analysis of facial movement synergies} using AUs as sources. We present two selection methods to analyze co-occurring activation patterns of AU6 and AU12 in the context of posed and spontaneous smiles.
\end{enumerate}

\section{Related work}
\subsection{EMG-based identification}
Compared to traditional EMG measurements, a reduced set of electrode positions has proven to yield high facial expression recognition rates of $87$\% accuracy for seven posed facial expressions, including sadness, anger, disgust, fear, happiness, surprise and neutral expressions. This subset includes electrodes placed on the \textit{Corrugator} and \textit{Frontalis} on the forehead; and ZM and \textit{Masseter} on the cheeks~\cite{Schultz2010}. Distal EMG has been used to identify different facial gestures by using different electrode configurations. Two EMG bipolar channels were placed on the \textit{Temporalis} muscle on each side of the face, and one placed on the \textit{Frontalis} muscle gave input to distinguish ten facial expressions. The achieved accuracy was $87$\% using a very fast versatile elliptic basis function neural network (VEBFNN)~\cite{Hamedi2013}. Although not all gestures were facial expressions of emotion, they did include symmetrical and asymmetrical smiling, raising eyebrows, and frowning. 

Distal EMG has been implemented on a wearable designed to keep four EMG channels attached to the sides of the face at eye level (Fig.~\ref{fig:wearable}). With this placement, it is possible to reliably measure smiles in different situations without obstructing facial movement~\cite{Perusquia-Hernandez2017,Funahashi2014}. This is possible because smile-related distal activity measured from the ZM is sufficiently large to be robust against non-affective facial movements such as chewing gum and biting~\cite{Oberman2007,VanBoxtel2010,Gruebler2014}. Hence, the information picked up by the four channels is used to approximate different sources of muscular activity using ICA~\cite{Comon1994}. The separated muscle activity contains components for muscles involved in generating smiles and can be used to identify these~\cite{Gruebler2014}. This approach can be used offline for fast and subtle spontaneous smile identification~\cite{Perusquia-Hernandez2017} and is possible even in real time~\cite{Takano2014}. Finally, this device has also been used to analyze spatio-temporal features of a smile by fitting envelopes to the EMG’s Independent Components (ICs), and later performing automatic peak detection on those envelopes~\cite{Perusquia-Hernandez2017a} with performance similar to that achieved by Computer Vision~\cite{Perusquia-Hernandez2019}. Furthermore, four EMG leads placed around the eyes in a Head-Mounted Display (HMD) have also been used successfully to distally identify facial expressions even when the face is covered by the device. Facial expressions of anger, happiness, fear, sadness, surprise, neutral, clenching, kissing, asymmetric smiles, and frowning were identified with $85$\% of accuracy~\cite{Cha2020}. Another recent work proposed the use of a thin sticker-like hemifacial 16 electrode array to paste on one side of the face and identify ten distinct facial building blocks (FBB) of different voluntary smiles. Their electrode approach is novel, robust against occlusion, and provides a higher density electrode array than that of the aforementioned arrangements. This enabled them to use ICA and clustering to define several FBB corresponding to a certain muscle~\cite{Inzelberg2020}. Nevertheless, they require electrode usage proximal to each muscle. This entails that a large sticker needs to be placed on the skin, obstructing spontaneous facial movement. Moreover, the physical connection of the electrode array enhances artifact cross-talk between electrodes. To eliminate such cross-talk, ICA was used and the resulting clusters were derived manually.

\begin{figure*}
  \centering
  \includegraphics[width=0.8\textwidth]{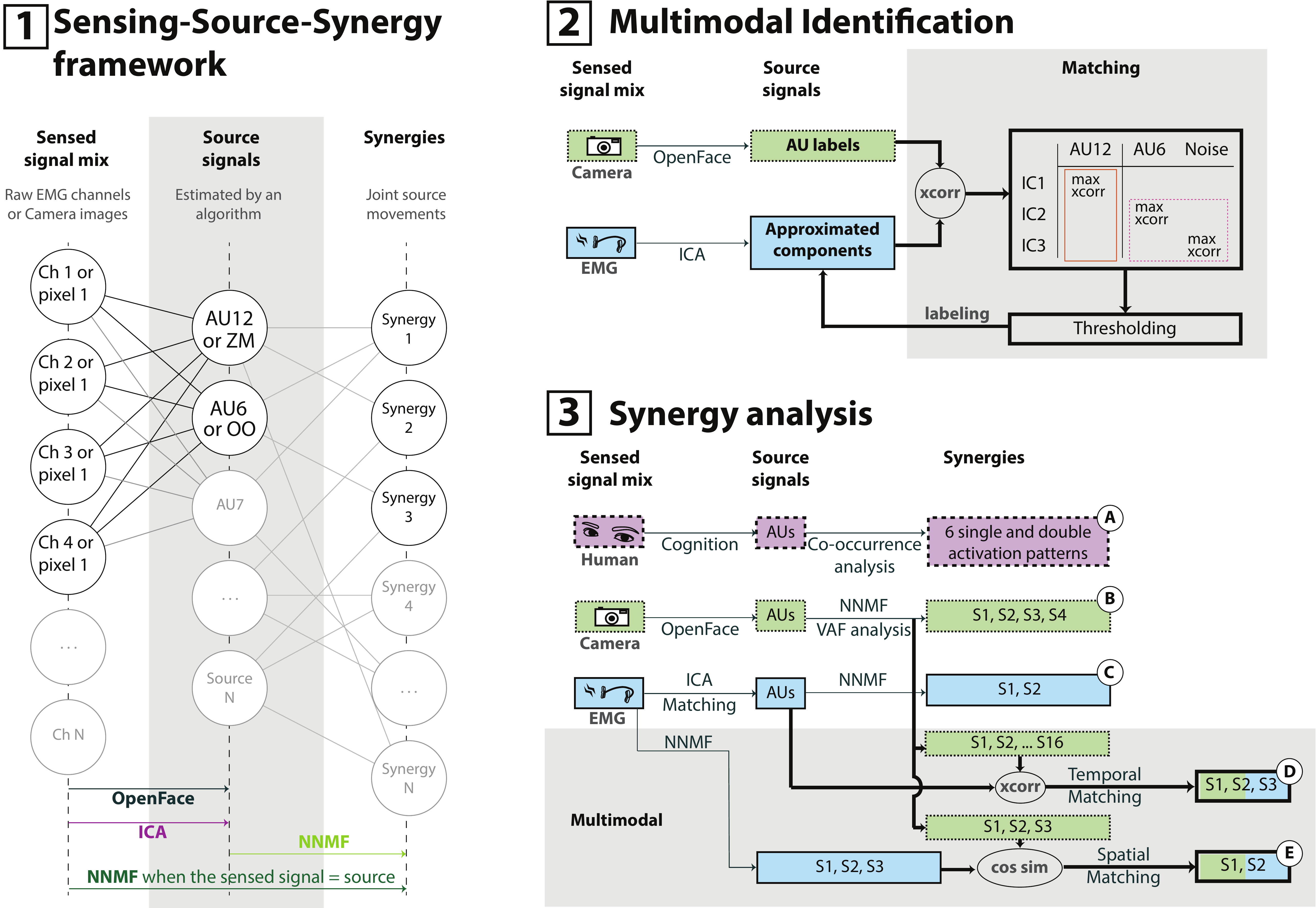}
  \caption{(1) Sensing-Source-Synergy framework. Sensors often read raw signals that are a mixture of signals of interest and other sources that can be considered noise. In other cases, the measured signals can be considered directly as the sources. We use OpenFace and ICA to derive movement source units. (2) Multimodal AU identification. AU labels are extracted from CV, and are used to assess facial expressions. Then the information derived from CV is used to identify the EMG components that correspond to each AU type. (3) Synergy analysis. Groups of sources moving together were analyzed by looking at the different sensing modalities independently or in combinations. (A) Six single and double activation patterns were proposed and identified from discrete AU labels. (B) Using NNMF and VAF methods, four AU synergies were found. (C) ICA was used to find the muscle sources and then NNMF was applied on the ICs to identify two synergies. (D) Temporal matching of each of the possible synergies with the ICs identified from EMG. (E) Spatial matching of the NNMF weights derived from EMG and four CV-labeled AUs selected from the VAF analysis.}
  \label{fig:framework}
\end{figure*}

\subsection{CV-based identification}
CV is the most widely used technique for identifying facial expressions~\cite{Bettadapura2012b}, even at the individual Action Units level. There are different approaches to extract relevant features for AU identification and intensity estimation. Among these, appearance-based, geometry-based, motion-based, and hybrid approaches. Several algorithms range between 0.45 and 0.57 F1 scores for occurrence detection and between 0.21 and 0.41 for intensity estimation~\cite{Martinez2017}. The OpenFace toolkit 2.0~\cite{Baltrusaitis2018} is a CV pipeline for facial and head behavior identification. Its behavior analysis pipeline includes landmark detection, head pose and eye gaze estimation, and facial action unit recognition. This algorithm detects AU 1, 2, 4, 5, 6, 9, 12, 15, 17, 20, 25, 26 with an average accuracy of 0.59 in a person-independent model. Moreover, the use of spatial patterns has been shown to achieve about 90\% accuracy in the task of distinguishing between posed and spontaneous smiles \cite{Wang2016}. Dynamic features based on lip and eye landmark movements have provided an identification accuracy up to 92.90\%~\cite{Dibeklioglu2015a}. Other algorithms using spatio-temporal features as identified by restricted Boltzmann machines have been able to achieve up to 97.34\% accuracy in distinguishing spontaneous vs. posed facial expressions~\cite{Yang2017}.

\subsection{Muscle synergies}
The concept of synergy in facial activity recognition has been used mainly to describe synergies between different sensors~\cite{Kostinger2012,HarisKhan2017,Henry2018}. However, muscle synergies refer to simultaneous muscle activation. Muscle synergies are typically expressed in the form of a spatial component (synergies) and a temporal component (activations). The spatial component describes the the grouping and ratio of muscles that are activating together for a given movement. The temporal component describes how each spatial component is activated in the time series. Two types of synergies have been currently proposed: "Synchronous synergies" assume there are no temporal delays between the different muscles forming the synergies (i.e. synergies are consistent throughout the movement) while the activation of these components change. On the other hand, "time-varying synergies" consider both the synergies to change~\cite{Tresch2009}. There is an ongoing debate on whether the Central Nervous System controls the activation of individual motor units, individual muscles, group of muscles, or kinematic and automatic features~\cite{Tresch2009}. This research is mainly done in the domain of motor control of wide and coordinated movements such as gait. Several researchers have used NNMF as a method to identify muscle synergies during posture and gait responses. This method is often used jointly with an analysis of Variance-Accounted-For (VAF) by each synergy component when reconstructing the original EMG signal~\cite{Torres-Oviedo2010,Torres-Oviedo2007,Tan2018,Tan2020}. In this method, the source signals are decomposed in as many components as there are degrees of freedom. The number of components that contain a VAF higher than a threshold are considered as the number of synergies contained in the group of sources.

\section{The Sensing-Source-Synergy framework}

We propose a framework to analyze sensed signals by estimating their sources and synergies.
Since AUs are closely related to individual muscle activity, we refer to them as ``sources" (Fig.~\ref{fig:framework}). Sources are facial movement units caused by a certain muscle. These individual sources often move in synchronous manner to form visible facial expressions such as smiles. A group of muscles, or a group of AUs, moving together are called synergies. Different transformations are necessary to go between sensed signals, source signals, and synergies. To go between a sensed signal mix to the source signals originating a movement, we can use ICA for EMG, and OpenFace for videos. Similarly, NNMF can be used to go from movement sources to synergy groups. A special case is when the sensed signal is very close to the source signal. For example, if EMG is measured directly from the muscle originating the muscle activity, we can assume equivalence between measurement and source. In such cases, NNMF can be applied directly to the sensed signal to obtain synergies.

\section{Data Set}
This data is a subset of the data generated in a previous study~\cite{Perusquia-Hernandez2019c}. Here only a brief description is provided for informative purposes.

\subsection{Participants}\label{PosSpont:ss:exp2part}

41 producers took part in the study (19 female, average age=25.03 years, SD=3.83). 

\subsection{Experiment design}
The experiment consisted of several blocks. All the producers completed all the experimental blocks in the same order. This was to keep the purpose of the experiment hidden during the spontaneous block.

\subsubsection{Spontaneous Block (S-B)} A positive affective state was induced using a 90~s humorous video. After the stimuli, a standardized scale assessing emotional experience was answered. Next, producers were asked to tag any facial expressions that they had made. 
\subsubsection{Posed Block (P-B)} Producers were requested to make similar smiles as they did in the S-B. However, this time, a 90~s slightly negative video was presented instead. Their instruction was: ``Please perform the smiles you video coded. This is for a contest. We are going to show the video we record to another person, who is unknown to you, and if she or he cannot guess what video you were watching, then you are a good actor. Please do your best to beat the evaluator". After watching the video and performing the task, they completed the same standardized scale assessing emotional experience. They were also asked to tag their own expressions.

\subsection{Measurements}\label{PosSpont:ss:exp2measure}

\begin{itemize}
    \item \textbf{Smile-reader.} Four channels total of distal facial EMG were measured from both sides of the face using dry-active electrodes (Biolog DL4000, S\&ME Inc) sampled at 1~kHz (Fig.~\ref{fig:wearable}). 
    \item \textbf{Video recordings.} A video of the producer’s facial expressions was recorded using a Canon Ivis 52 camera at 30~FPS.
    \item \textbf{Self video coding.} The producers tagged the onset and offset of their own facial expressions using Dartfish 3.2. They labeled each expression as spontaneous or posed, and indicated whether or not it was a smile. 
    \item \textbf{Third person video coding.} Two independent raters labeled the videos with Dartfish 3.2. They coded for the start frame and the duration of every smile, and AUs 1, 2, 4, 5, 6, 9, 10, 12, 14, 15, 17, 18, 25, 26 and 28.
\end{itemize}

\section{Data analysis}
We introduced a novel framework to estimate muscle movement sources from distal EMG. Using this framework, we analyzed facial muscle synergies in an automatic manner. Two types of data analysis were performed (Fig.~\ref{fig:framework}). The first type is multimodal identification, which aims to identify different sources or muscle groups from the recorded EMG; and assesses their similitude to AUs detected using CV. We rely on feature engineering using blind-source separation, cross-correlation, and dimension reduction. This is, we are using feature engineering applied on a continuous-time series to transform the data into more relevant information that can be thresholded by its SD. The second one is synergy analysis, aimed to identify the spatial and temporal structures of the facial expressions present in the data. The synergy analyses were conducted both on single modalities and at the multimodal level. Both types of analysis used the same type of EMG pre-processing. 

\textbf{EMG Pre-processing.} The four EMG channels were first passed through a custom Hanning window with a ramp time of 0.5~s to avoid introduction of artificial frequencies by the filtering at the start and the end of the signal. Afterwards, the signals were (1) linear detrended, (2) transformed to have zero mean and one standard deviation, (3) band-pass filtered from 15 to 490~Hz, (4) rectified, and (5) low-pass filtered at 4~Hz. 
    
\textbf{CV-based AU labeling using OpenFace.} The Facial Behavior Analysis Toolkit OpenFace 2.0 was used to identify several facial features including AUs. AU identification is given both as a continuous output or intensity rating; and a binary output indicating AU presence. The intensity and presence predictors have been trained separately and on slightly different datasets, which means that they are not always consistent~\cite{Baltrusaitis2018}. In this work, we choose to use the continuous or the binary rating depending on the requirements of our algorithm. The binary CV AU labels are extracted and upsampled, as well as the continuous labels, from 30~Hz to 1~kHz to match the EMG sampling frequency.
    
\textbf{Blind-source separation.} ICA~\cite{Hyvarinen2000}, was used to automatically estimate different muscle activity sources. The wearable used to collect the data has four channels. Thus, we set the number of decomposed components to three. 
    
\textbf{Synergy identification.} NNMF~\cite{Lee1999} is a dimensional reduction method aimed to uncover synchronized muscle movements. We expect it to be able to identify source activity happening at the same time, either from CV-AUs or EMG. If AU6 and AU12 happen at different times, they should be categorized as belonging to different synergies. The used dataset contains both posed and spontaneous smiles. Thus, we hypothesize that NNMF will be able to identify whether AU6 and AU12 belong to the same synergy or not. The number of synergies found primarily depends on the number of available sources. In the case of CV, 17 AUs were identified. Hence, the degrees of freedom were a maximum of 16. In the case of EMG, the degrees of freedom is three when applying NNMF on the raw measurements, and two when applied on the estimated ICs. Fig.~\ref{fig:framework}-3 shows in detail the processing followed.
    
\textbf{Multimodal identification with component matching to CV-generated labels.} A matching method was used to assess similarity between EMG components and CV-based labels (Fig.~\ref{fig:framework}-2). We assume the EMG signal to contain AU6, AU12 and noise. Noise is defined as electrical interference as well as other muscular sources. First, we calculated the cross-correlation of the three ICA components; the continuous AU6, AU12 OpenFace CV-labels; and an uniformly distributed random noise distribution. Since AU12 stems from the large and strong ZM muscle, the index of the maximum correlation is chosen to correspond to AU12. The other two ICs get assigned to be AU6 and noise in order of maximum correlation value. Afterwards, a threshold method was used to determine active samples. Further smoothing is applied on the individual ICs by means of a first order Savitzky-Golay filter with length 301. An initial period of $\approx 1~s$ or 30 samples of the IC is used to calculate the baseline signal average and standard deviation. The whole signal then is turned into a binary vector where samples that cross the threshold of $\overline{m} + k\sigma$ with $k=2$ are set to one. The values set to one are thought to correspond to activity of the respecting AU assigned to the IC during the process of AU identification. 
In this process, we were careful not to use the ground truth labels (Human-coding) as input for our algorithm. The human labels were only used to assess how good the EMG threshold method is approximating visible activity tagged by the human coders. 
    
\textbf{Synergy Analysis.} In some contexts, AUs might appear together more often. This might be the case for the co-occurrence of AU6 and AU12 in spontaneous smiles, should the Duchenne marker truly be a marker of smile spontaneity. In contrast, posed smiles would be characterized by less synchronized AU6 and AU12 activity. Hence, several co-occurring activation were analyzed using different modalities (Fig.~\ref{fig:framework}-3-A). We propose six activation patterns: (1) AU6 only; (2) AU12 only; (3) AU12 inside AU6; (4) AU6 inside AU12; (5) AU12 before AU6; and (6) AU6 before AU12. An algorithm was designed to detect these from binary labels. These labels can be generated by human coding, CV, EMG or a combination of them. This method relies on subtracting the labels of one AU from the other, and then identifying the differences within each block for each activation pattern.
    
    
NNMF was used to extract muscle synergies from EMG, and AU synergies from the continuous CV-labels. The total variance accounted for (VAF) was used as a metric to determine the ideal number of synergies~\cite{Tan2018}. First, we used the AUs derived from CV (Fig.~\ref{fig:framework}-3-B), and decomposed them into NNMF components between 1 and 16 per experimental block. Afterwards, the ideal number of synergies was determined by ensuring that the number of synergies would reconstruct the original signal with less than 15\% error for all participants in both posed and spontaneous blocks. 
Furthermore, we compared the EMG-based and CV-based synergy detection to match them across modalities. The matching was done in the temporal domain using cross-correlation (Fig.~\ref{fig:framework}-3-D), and in the spatial domain using cosine similarity to sort and match the cross-modal NNMF component weights (Fig.~\ref{fig:framework}-3-E). To use cosine similarity, it is necessary to have components with equal number of weights in both modalities. Thus, we decomposed three synergies from only four smile-related CV-labeled AUs (AU 6, 7, 10 and 12). This was to match the four EMG channels. Nevertheless, we would prefer to do a cascading transformation by first determining the sources using ICA, and then applying NNMF to the ICs (Fig.~\ref{fig:framework}-3-C). However, due to the limited number of EMG channels and the evidence from the multimodal identification that AU12 and AU6 are strong in the raw signal; we considered the EMG signal to be equivalent to the sources only for this analysis (Fig.~\ref{fig:framework}-3-E). Given this, the spatial matching would be an alternative to the multimodal identification described in Fig.~\ref{fig:framework}-2.
    
\textbf{Delay between raw EMG and CV-based AU detection.} The delay between the EMG signals and the CV-based AU detection was calculated by looking at the lag at which the maximum cross-correlation between EMG channels and AU labels appeared. A similar method was used to calculate the delay between AU6 and AU12 from the CV-based binary labels.
    
\textbf{Agreement with human coders.} Human-coded labels, CV-labels, and EMG-labels were transformed to a matching sampling rate. Then the agreement between different measurements was calculated using Cohen's Kappa~\cite{Cohen1968}. Additionally, we report accuracy, precision, and recall. The advantage of using Cohen's Kappa is that it penalizes for the larger amount of no AU samples in the set, given that participants did not smile all the time.

\section{Results}

\subsection{Delay between CV-EMG signals}
The delay between raw EMG and CV-based labels was 374~ms in average (median~=~450~ms, SD~=~366~ms, Fig.~\ref{fig:processing}).

\subsection{Action Unit identification}
\subsubsection{Ground truth}
The inter-coder agreement was a Cohen's Kappa of 0.78 for AU06 and 0.84 for AU12. For further processing, a single human-coded label was set to active when either of the coders thought there was an AU. The agreement between CV and the human labeling was a Cohen's Kappa of 0.42 for AU06 and 0.43 for AU12.

\subsubsection{Multimodal identification}
\begin{figure}
    \centering
    \includegraphics[width=0.50\textwidth]{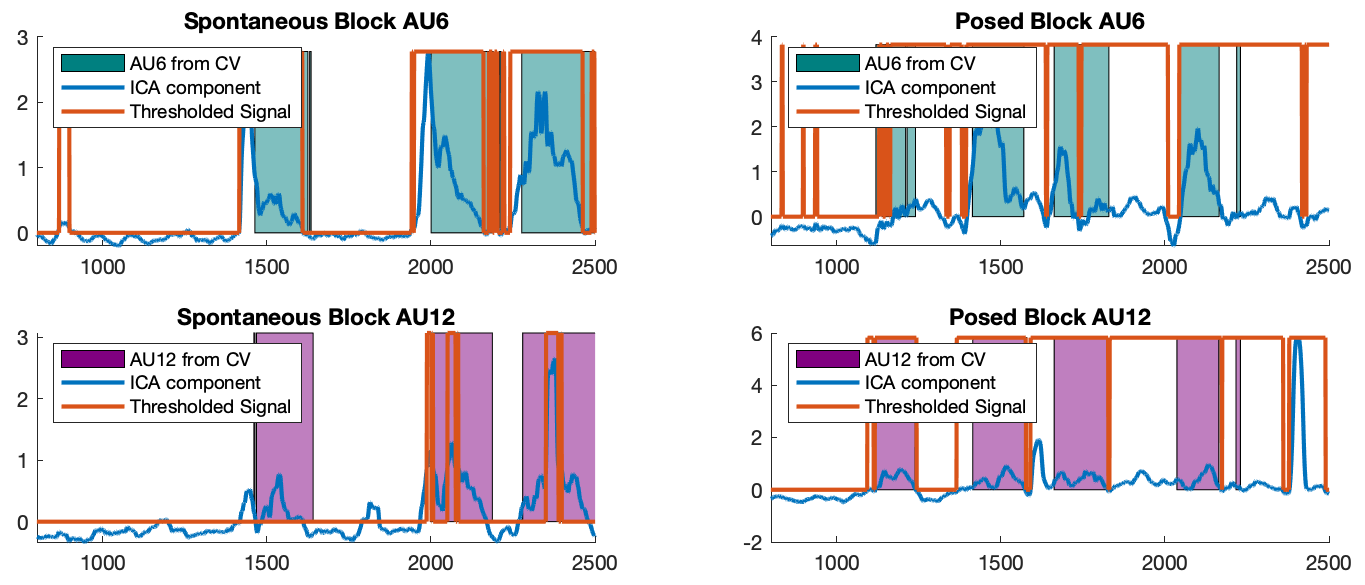}
    \caption{The threshold method applied to the ICA components identified as AU6 and AU12 for a block with spontaneous smiles (left) and posed smiles (right). The ICA components are show as blue lines. 
    }
    \label{fig:threshold_result_example}
\end{figure}

The Cohen's kappa between Human-coded labels and EMG-based labels was 0.49 for AU6 and 0.53 for AU12. Furthermore, the accuracy reached 81\% for AU6 and 82\% for AU12 (Tab.~\ref{tab:agreement}). 
Although the correlation between CV-based labels and EMG-based labels is an important step for the selection of the EMG components, the selected EMG-based labels might not coincide perfectly with the CV-based labels (Fig.~\ref{fig:threshold_result_example}).

\begin{table}[!t]
\renewcommand{\arraystretch}{1.3}
\caption{Agreement of AU6 and AU12 between human-coded and CV-based and between human-coded and CV-EMG detected AUs.}
\label{tab:agreement}
\centering
\begin{tabular}{|ll|lll|}
\hline
     &                    & Coder1     & Coders  & Coders \\
     &                    & vs. Coder2 & vs. CV & vs. EMG \\
\hline
AU6  & Agreement $\kappa$ & 0.78 & 0.42 & 0.49 \\ 
     & Accuracy           & -    & 0.83 & 0.81 \\
     & Precision          & -    & 0.93 & 0.62 \\
     & Recall             & -    & 0.34 & 0.60 \\
\hline
AU12 & Agreement $\kappa$ & 0.84 & 0.43 & 0.53 \\ 
     & Accuracy           & -    & 0.83 & 0.82 \\
     & Precision          & -    & 0.90 & 0.63 \\
     & Recall             & -    & 0.35 & 0.68 \\
\hline
\end{tabular}
\end{table}

\subsection{Synergy analysis}

\subsubsection{Co-occurrence analysis} 
\begin{figure}
    \centering
    \includegraphics[width=0.35\textwidth]{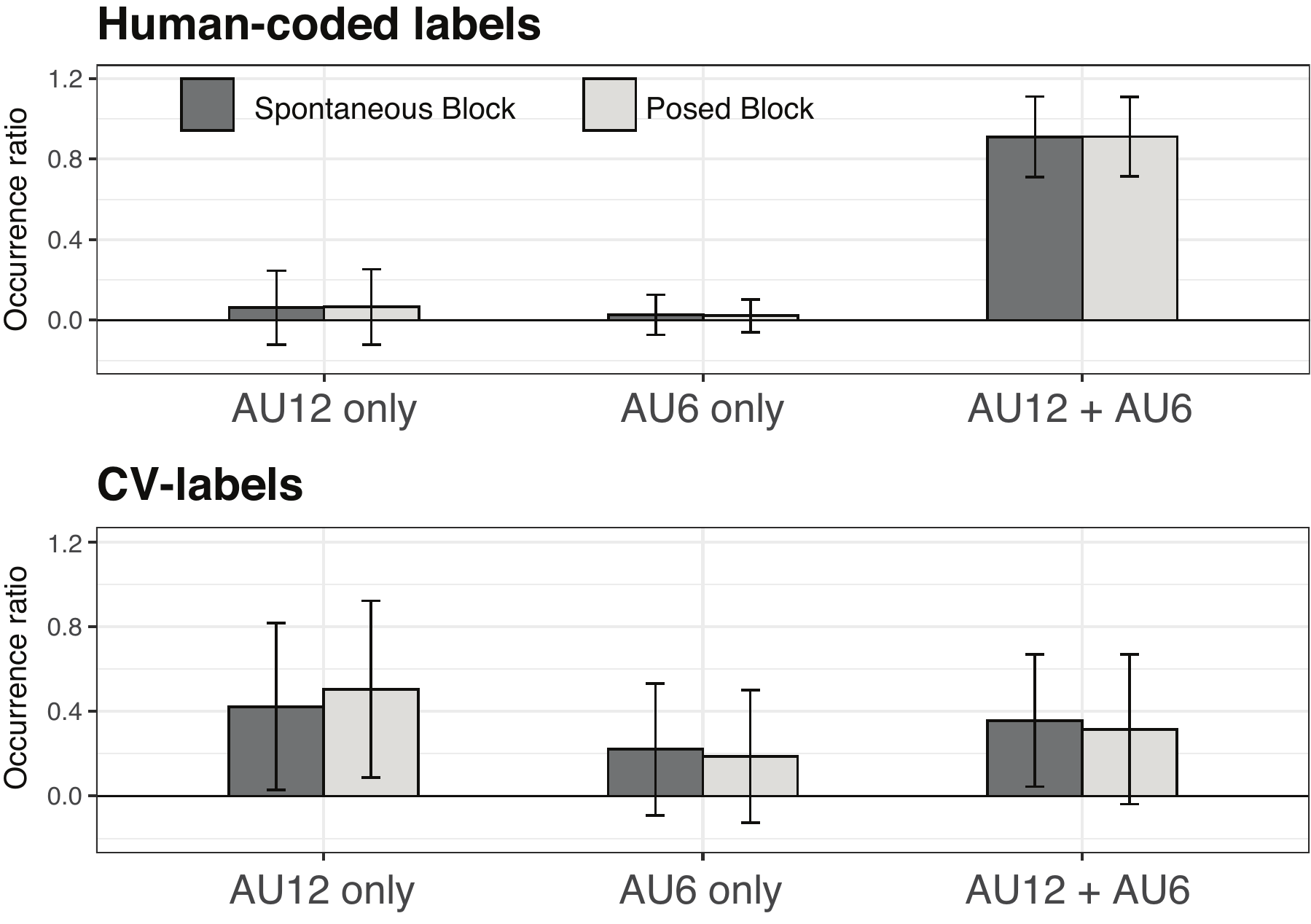}
    \caption{According to human coders, AU6 and AU12 co-occur most of the time. This differs from the labels according to OpenFace as expected from the low agreement between computer vision and human labels.}
    \label{fig:auBlockResults}
\end{figure}

\begin{figure}
    \centering
    \includegraphics[width=0.48\textwidth]{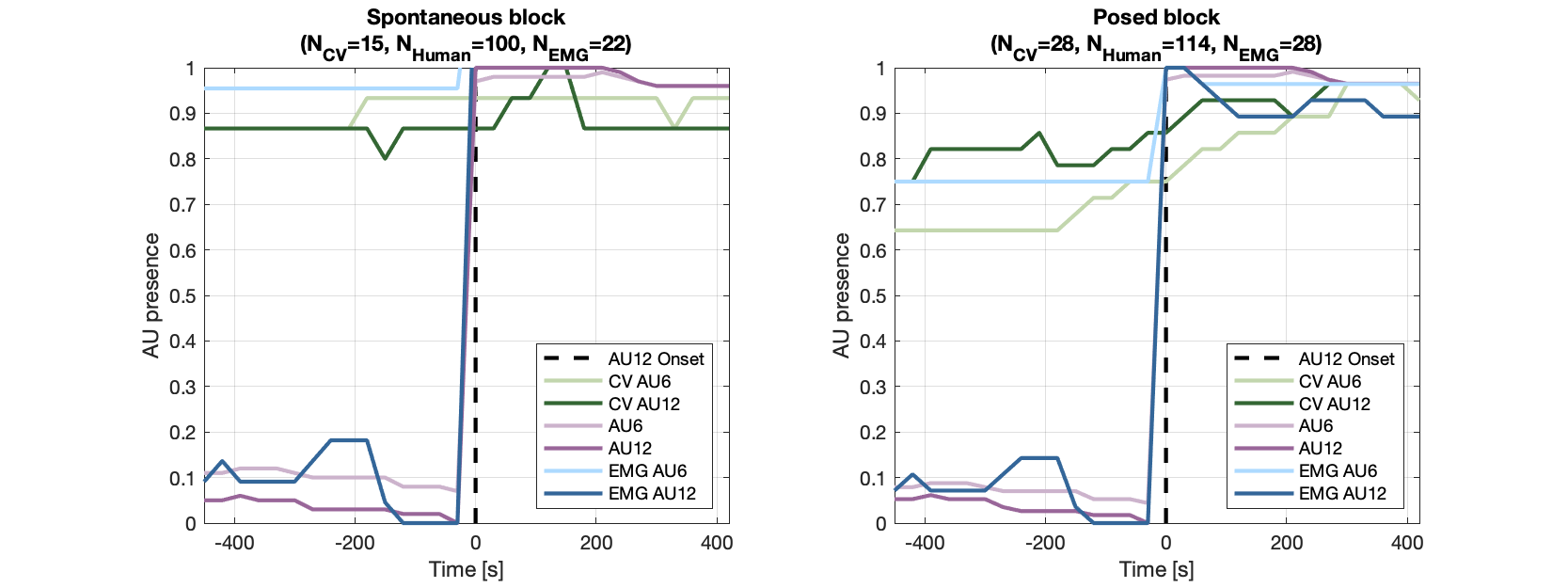}
    \caption{Frame-by-frame comparison of AU6 with respect to the onset of AU12 from human coded AUs, CV extracted AUs and AUs labeled with our method. The left figure shows results from spontaneous blocks, while the right figure shows the results from posed blocks. Human coded AUs are depicted in pink, CV extracted AUs in green and AUs from our method in blue.}
    \label{fig:au6_au12_temporal_comparison}
\end{figure}

To assess the co-occurrence frequency between AU6 and AU12, we evaluated how much agreement AU6 and AU12 labels have with each other. High agreement indicates that both AUs appear at the same time. For the human labeled data we see a Cohen's Kappa of $0.84$ (Tab.~\ref{tab:auto_agreement}). Similarly, the CV-based labels show a Cohen's Kappa of $0.62$. Finally, the AUs detected from EMG show a Cohen's Kappa of $0.34$. A Wilcoxon rank sum test on the AU co-occurence pattern analysis (Fig.~\ref{fig:auBlockResults}) showed that AU6 and AU12 co-occur more often than not (W~=~6519.5, p~\textless~.01), regardless of the experimental block (W~=~3362, p~\textgreater~.5). Furthermore, AU6 tends to start before AU12 most of the time (W~=~7385, p~\textless~.01). An analogous analysis of CV vision data yielded no significant differences between posed and spontaneous blocks (W~=~1390, p~\textgreater~.5), but showed opposite results regarding the order of occurrence between AU6 and AU12. According to CV, AU12 occurred before AU6 more often (W~=~0.71, p~\textless~.001). To better understand the temporal relationship between both AUs, we performed a frame-by-frame comparison of AU6 and AU12 activation from human coded, with CV extracted AUs and AUs labeled with our method (Fig.~\ref{fig:au6_au12_temporal_comparison}). Each frame was selected with respect to the onset of AU12. A frame started $0.5$~seconds before the onset and had a duration of one~second. Whereas human coders rated an almost perfect co-occurrence, EMG and CV coded a higher probability of AU6 being active before AU12 onset.

\begin{table}[!t]
\renewcommand{\arraystretch}{1.3}
\caption{Agreement of co-occurrence between AU6 and AU12 for human-coded, CV-based and from EMG detected AUs.}
\label{tab:auto_agreement}
\centering
\begin{tabular}{|l|l|l|l|l|}
\hline
	& Agreement $\kappa$ & Accuracy & Precision & Recall \\
\hline
Human   & 0.84 & 0.96 & 0.77 & 0.77 \\
CV      & 0.62 & 0.91 & 0.57 & 0.57 \\
EMG     & 0.34 & 0.75 & 0.48 & 0.48 \\
\hline
\end{tabular}
\end{table}

\subsubsection{NNMF VAF Analysis on CV-based AU continuous labels} 
\begin{figure}
    \centering
    \includegraphics[width=0.45\textwidth]{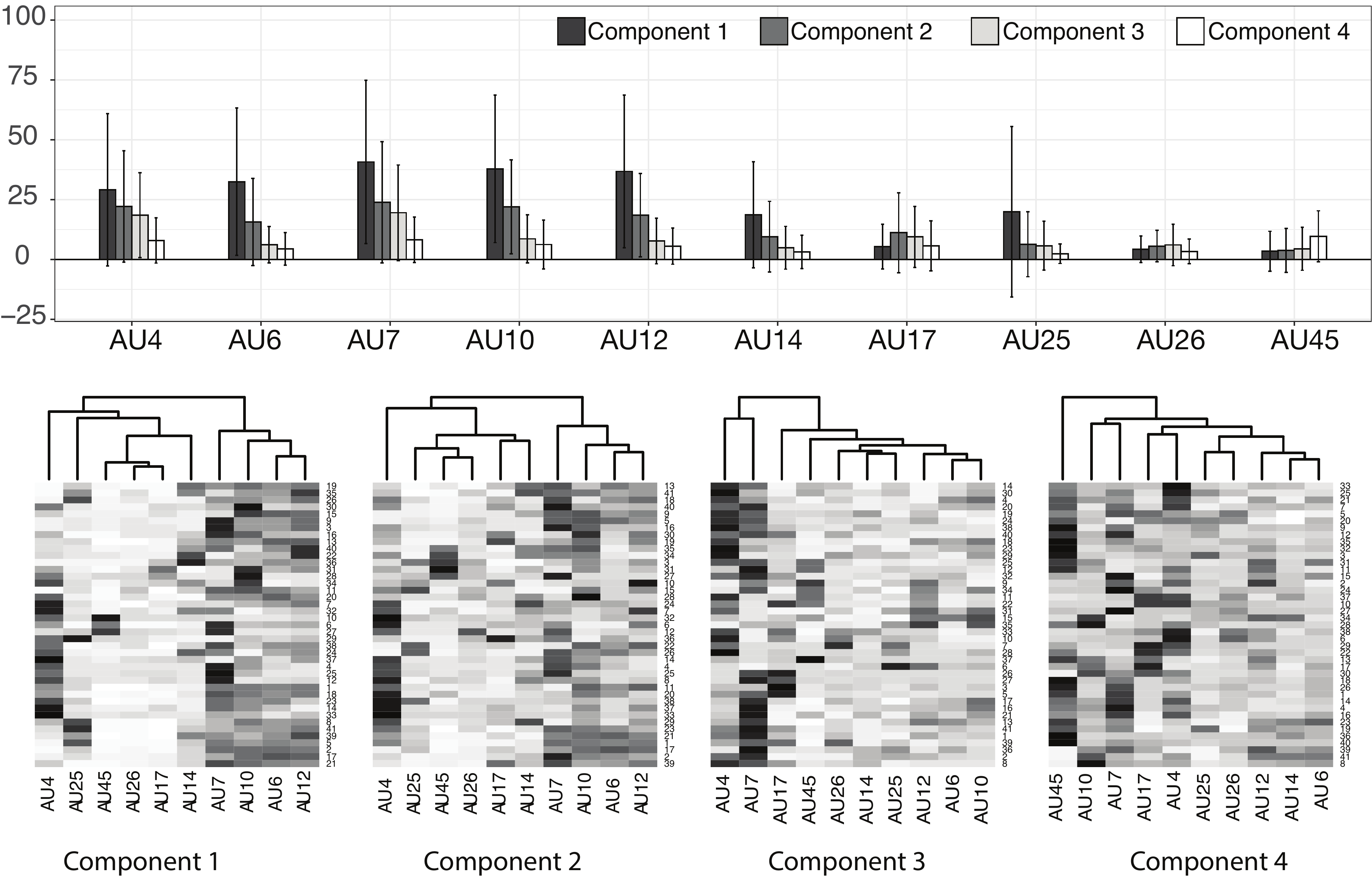}
    \caption{Weight averages per AU for the four components selected using the NNMF VAF as criteria. AU6 and 12 are grouped in components 1 and 2.}
    \label{fig:B_synergyWeightAverage}
\end{figure}

After iteratively decomposing the synergies from CV-AU labels using NNMF, the VAF analyses showed that four synergies account for 85\% or more of the variance in both blocks for all participants. 
A Wilcoxon rank sum test showed no differences in weights between spontaneous and posed blocks (W~=~3968700, p~\textgreater~.05). A closer look to the weights of those four synergies showed that not all AUs have high weights. Therefore, we selected the AUs whose weights contributed more to the four synergies. The selected AUs are 4, 6, 7, 10, 12, 14, 17, 25, 26, 45 (Fig.~\ref{fig:B_synergyWeightAverage}). The resulting weights were clustered per participant and AU within each synergy. AU6 and 12 were grouped together in the first two components, which accounted for most of the variance. AU4 was often present, but not clustered with other AUs. In the third component, AU6 was clustered with AU10. Finally, blinks (AU45) emerged in the last component.


\subsubsection{ICA-NNMF source-synergy analyis} 
\begin{figure}
    \centering
    \includegraphics[width=0.3\textwidth]{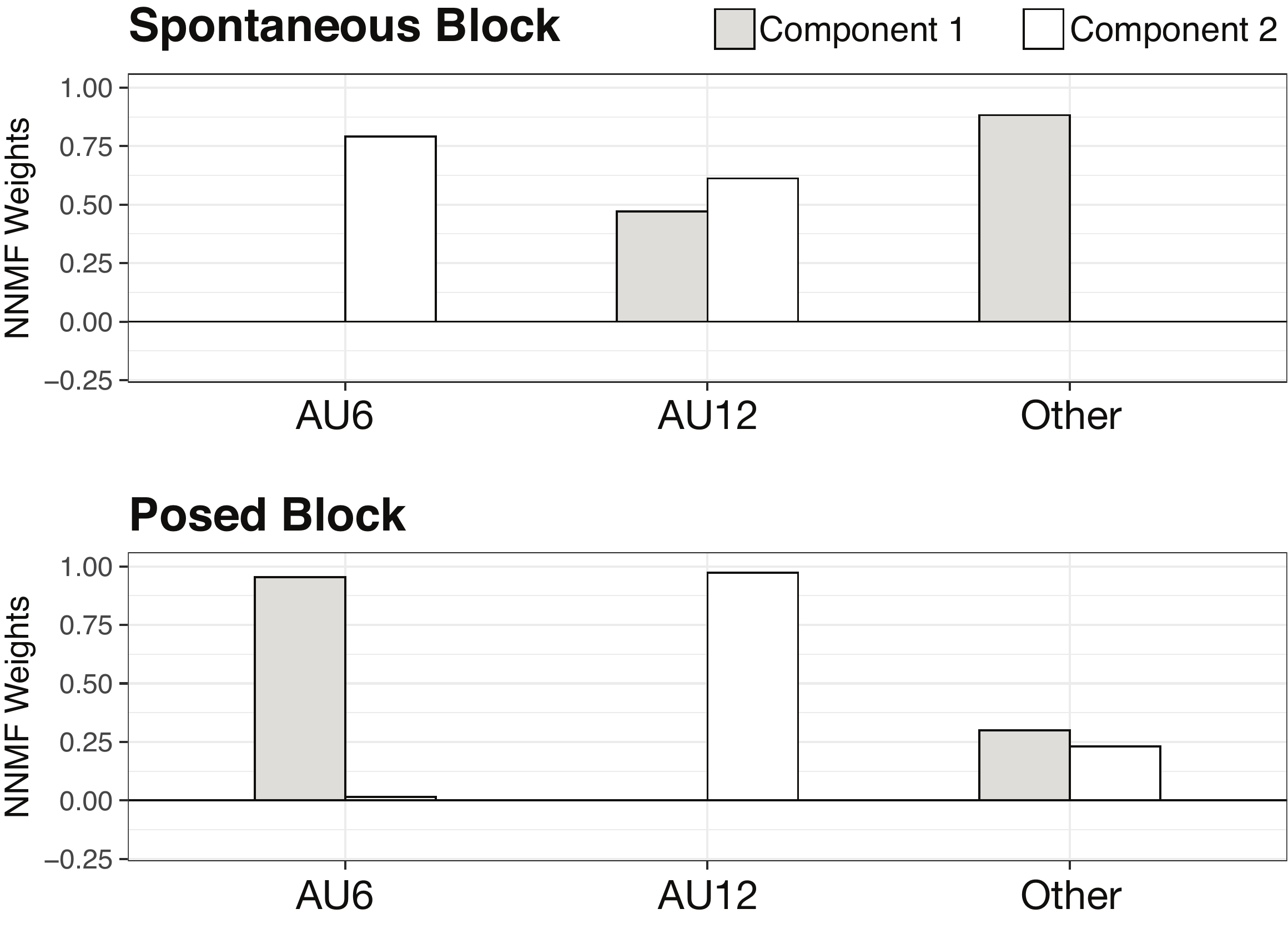}
    \caption{AU synergies identified per experimental block and NNMF-components. The horizontal axis shows the AUs detected by our matching algorithm. The vertical axis show the component weights per AU.}
    \label{fig:cascading}
\end{figure}

Since Distal EMG is not measured close to the movement originating muscle, we proposed to use ICA first to approximate the movement sources before applying NNMF. After ICA, three sources were approximated and labelled using our matching algorithm. Next, only two NNMF components can be approximated due to the lose of one degree of freedom after ICA. The weights for the two resulting synergy components are shown in Fig.~\ref{fig:cascading}. Interestingly, the difference between experimental blocks of the synergy weights of AU6 and AU12 was significant according to a Wilcoxon rank sum test (W~=~9886.5, p~\textless~.005). In the spontaneous block, one synergy is comprised of AU12 and others, and the second synergy is the joint activity of AU6 and AU12. On the contrary, the posed block yielded to two distinct synergies, one corresponding to AU6, and the other to AU12, which suggests that both facial movements are jointly executed only during spontaneous smiles. Finally, the "Other" label is comprised of any other AUs detected by the EMG electrodes.

\subsubsection{Temporal matching} 

\begin{figure}
    \centering
    \includegraphics[width=0.47\textwidth]{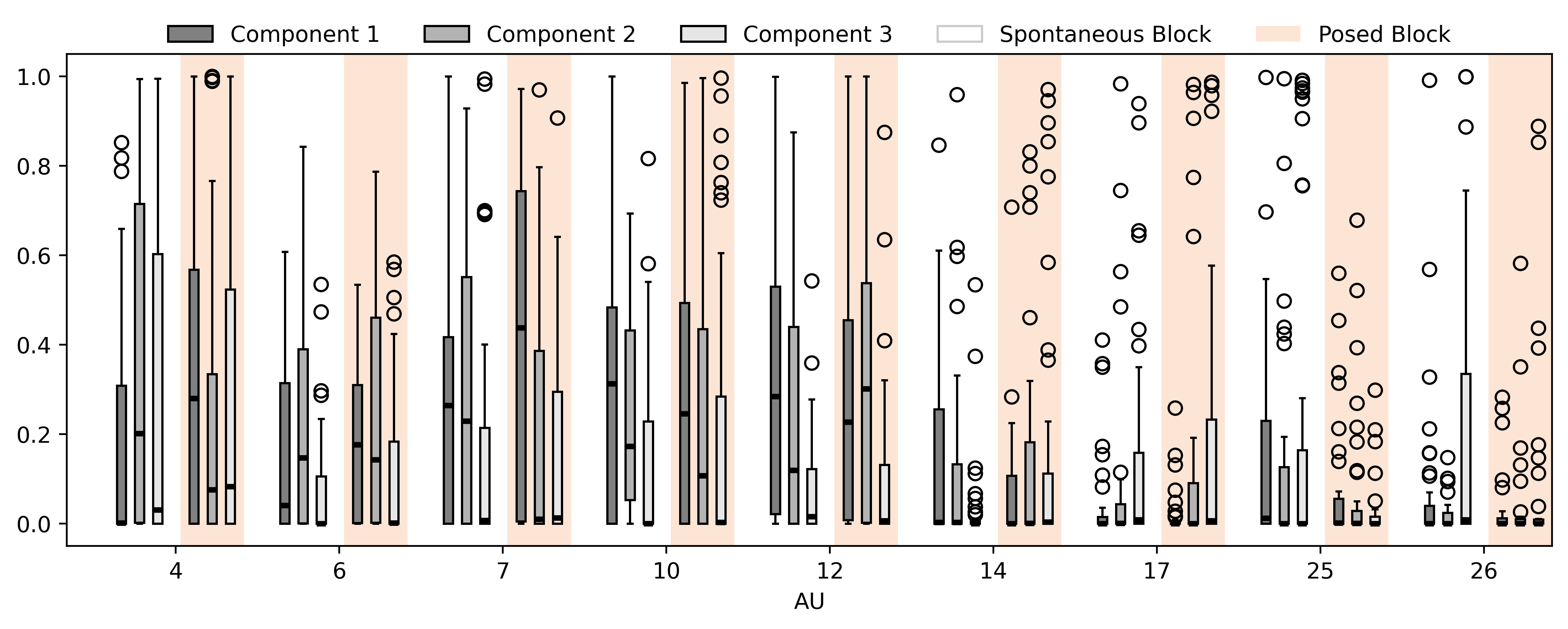}
    \caption{CV labeled AU weights after temporal matching with EMG components per experimental block. Components are grouped along the horizontal axis in AUs detected by OpenFace and experimental block. Each bar corresponds to the weight of a CV components with maximal correlation to the corresponding emg component.}
    \label{fig:chopchop}
\end{figure}

\begin{figure}
    \centering
    \includegraphics[width=0.47\textwidth]{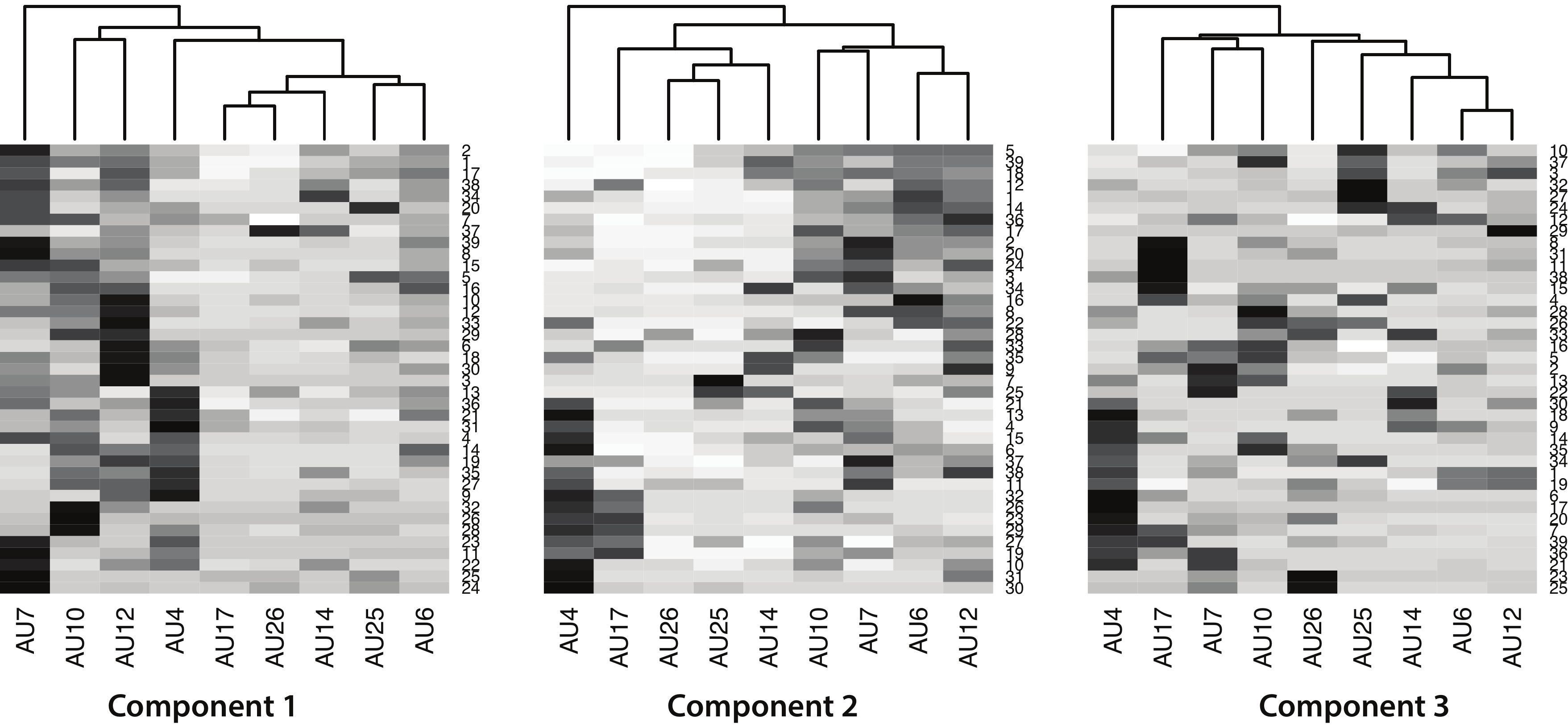}
    \caption{CV synergy cross-correlated with EMG IC Matching for selected AUs. Spontaneous and posed blocks were averaged given the lack of significant differences.}
    \label{fig:DtemporalMatching}
\end{figure}

Fig.~\ref{fig:chopchop} shows the CV-derived components per AU that are most correlated to the EMG identified AU activations per experimental block. A Wilcoxon rank sum test showed no significant difference between posed and spontaneous experimental blocks (W~=~3968700, p~\textgreater~.05). Fig.~\ref{fig:DtemporalMatching} shows a heatmap depicting the synergy weights per AU and participants. The AUs with weight loads close to zero were excluded. Since no differences were found between experimental blocks, the weights among the two blocks are averaged. AU6 and 12 were clustered together in two components, whilst in the first AU6 was grouped with AU25, suggesting that the participants smiled with lips apart. Furthermore, AU12 and AU10 were clustered together, which might be due to a confusion of the OpenFace algorithm.


\subsubsection{Spatial matching} 

\begin{figure}
    \centering
    \includegraphics[width=0.45\textwidth]{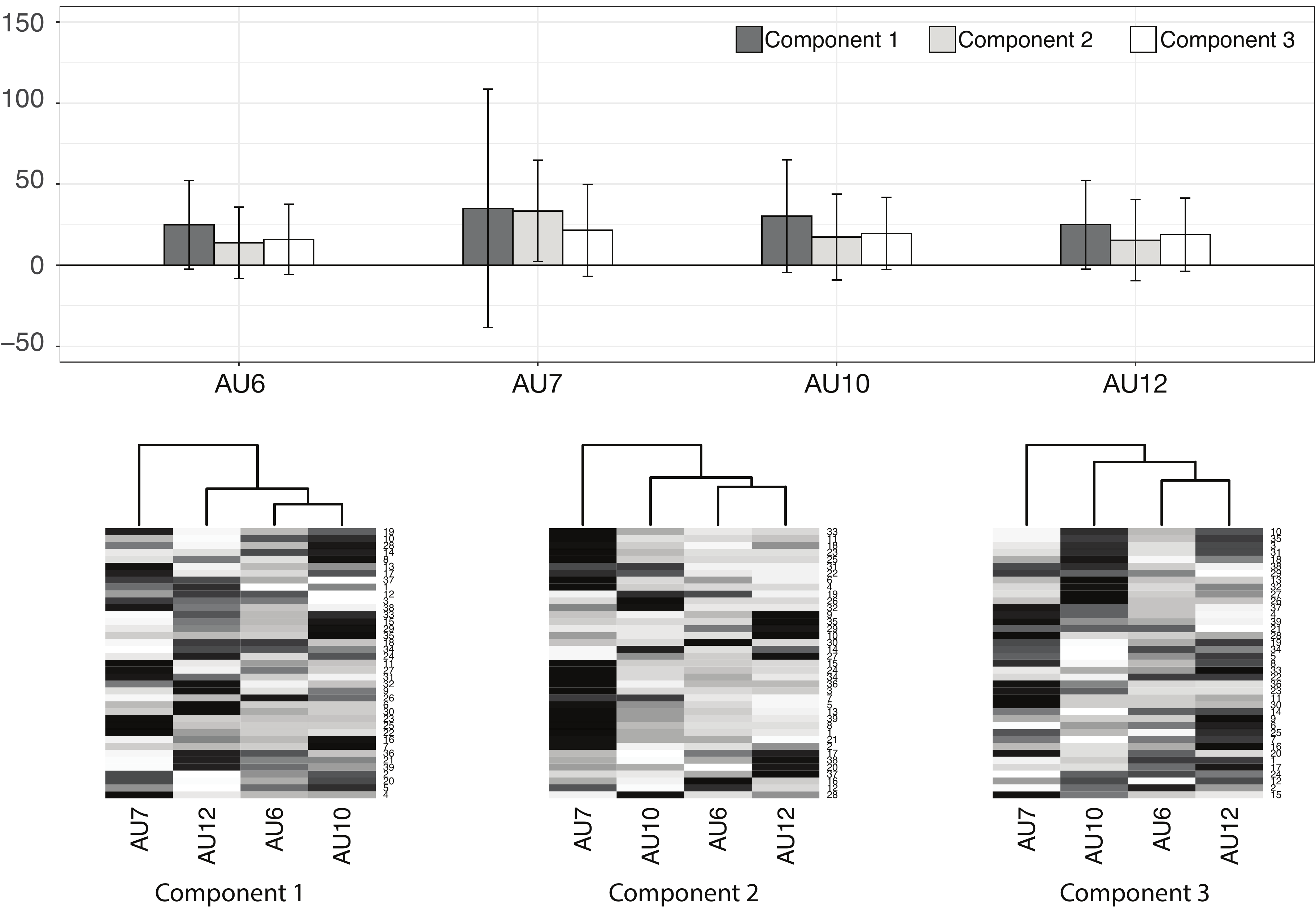}
    \caption{Four smile-related AUs were decomposed in three synergies for CV and EMG independently. The top plot shows the average weight loads per AU. The heatmaps below show the weight loads per participant and action unit per component. From the heatmaps, AU6 and AU12 were clustered together in components 2 and 3.}
    \label{fig:EspatialMatching}
\end{figure}

To match the dimensions of the EMG signal, which has four channels, four AUs were selected. These were AU6, AU7, AU10, and AU12. These are the AUs with the largest weight from previous analyses that are smile-related. Given the high occurrence of AU7 and AU10, we hypothesized that OpenFace might be confusing them with AU6 and AU12 respectively. A Wilcoxon rank sum test showed no differences in weights between spontaneous and posed blocks(W~=~27872, p~\textgreater~.05). Figure~\ref{fig:EspatialMatching} shows the synergy weights per AU. We can observe that AU6 and AU12 are clustered together in components 2 and 3. In component 1, AU6 is grouped with AU10 for most participants.

\section{Discussion}
We observed co-varying activation patterns of pre-processed EMG, and continuous AUs extracted from CV. There was a delay between CV AU activation and EMG activation, with EMG activation leading by 374~ms. This was expected as EMG originates skin displacement. This delay was larger than that observed from proximal EMG measurements (average of 230~ms)~\cite{Cohn2004}.

We showed that AU6 and AU12 can be detected from a multimodal algorithm that labels EMG signatures estimated with ICA with labels generated automatically with CV. These EMG signatures do not always correspond to those dictated by the CV algorithm during the calibration. However, they yield to a slightly higher agreement than if CV was used alone. The main reason why the accuracy of the CV labels did not constrain the accuracy of the EMG-derived labels is that CV is used only for initial identification of the EMG components that are more correlated to each AU. Afterwards, the EMG-based identification is independent of the OpenFace labels.

We proposed the framework of \textit{Sensing-Source-Synergies}. We distinguish the measurements made from the sources of the movement, and we use those sources to estimate synergies. We suggested to use ICA to search for activity from different muscle sources when measuring with EMG. On the other hand, NNMF attempts to find joint activation of different muscles. Several analysis pipelines in different sensing modalities were suggested to quantify the synergies between multiple sources. The human-coded co-occurring blocks of AU6 and AU12 showed that the Duchenne marker appeared simultaneously to the lip movements about 90\% of the time, independently of the experimental block. Interestingly, AU6 leading AU12 occurred in more cases than the opposite. A similar analysis on the binary CV-based labels showed a lesser percentage of simultaneous activation, followed by an even less simultaneous activation depicted by the EMG labels. This is probably because the automatic algorithms tend to detect gaps in between AU6 activation when human coders do not report so. Nevertheless, in the cases where the Duchenne marker and the lip movent co-occurred, AU6 was active before AU12 above 80\% of the time in the spontaneous block; and above 60\% of the time in the posed block. The agreement of AU6 and AU12 labels per modality was the highest for human-coding, followed by CV and EMG. Probably EMG had the least co-occurence agreement because of its higher temporal resolution, and, most importantly, because muscle activations lead visible activity. This suggests that indeed, muscle synergies do not perfectly match visible facial activity.

To better interpret the amount of AU synergies, NNMF was also performed on the AU labels derived from CV, and a VAF metric was used to determine the ideal number of synergies. The results suggested four synergies. The weights suggested that not all AUs contributed to these synergies, and no differences between posed and spontaneous experimental blocks was found. Another method proposed to select synergies was temporal matching with the EMG IC components. This method yielded three synergies with similar characteristics to the ones selected with the VAF criteria. Also, no significant differences were found between posed and spontaneous blocks. This entails that even though AU6 and AU12 weights are often clustered together within the selected synergies, there were no differences between posed and spontaneous smiles. According to these analyses, participants also displayed other AUs, notably AU4. This was expected, especially in the posed block where there might be a conflict between what the participants felt and the happiness they were asked to convey with a smile. It does not mean that those AUs are part of a smile, but that they were also present in the block. This suggested that antagonistic AU synergies might also occur, as to try to inhibit or mask facial expressions other than the intended one. Furthermore, we observed that OpenFace often lead to high weights in AUs that are similar to AU 6 and 12. For example, AU6 might be confused with AU7, and AU12 with AU10. Thus, when we had to further reduce the AUs to match to the EMG results, we chose these. The results showed that AU6 and 12 are often grouped together, and in one case, AU6 was grouped with AU10. Thus, whilst CV-based identification has a higher spatial resolution to identify several AUs, it still confuses several labels. Additionally, NNMF was used to identify joint muscle activity directly from the pre-processed EMG in an spatial matching algorithm. The results were similar to those found by applying NNMF on CV AUs only. This is in line with the hypothesis that NNMF considers the ZM and the OO to move as one single synergy. A closer look into the NNMF results suggested that the aforementioned muscles move in a single synergy picked up by the lower electrodes of the wearable. Thus, probably the synergy is dominated by the ZM. This is in line with results showing the significant strength of the ZM when compared to other muscles, and it might be related to muscle length~\cite{VanBoxtel2010}. To reiterate, our proposed method is using OpenFace as a rough guide for EMG labeling. Our algorithm does not need the OpenFace to be perfect, as we consider only the dynamics of the facial activations as input for the cross-correlation. Since the EMG data contains the movement sources measured distally, and ICA estimates them blindly, CV is only used for (soft-) labeling. In later stages, the algorithm is completely independent from the CV accuracy. Evidence of this is the superior performance of EMG with respect of CV with regards to the AU detection. In other words, the performance of the OpenFace algorithm did not constrain our overall detection accuracy.

We also used the results of our proposed multimodal identification of AUs using EMG, as an input to the NNMF algorithm. Although the number of synergies that can be identified using this method is only two, the results were surprising. We had hypothesized that a challenge for ICA might be to disentangle multiple AUs from the mixed EMG given the joint activation of the ZM and the OO. However, the results showed that ICA actually boosted the NNMF synergy detection by transforming the data to the source space first. Only in this case, we found a strong difference between posed and spontaneous smile AU synergies.
In posed smiles AU6 and 12 seem to operate independently, whereas a joint movement of AU 6 and 12 was observed in spontaneous smiles. This is somehow in line with the Duchenne marker hypothesis. Even though the appearance of the Duchenne marker can be simulated voluntarily, the underlying muscle synergies are distinct. The success of EMG to recognize subtle differences might be that (1) with our matching algorithm EMG already contains information from the CV-based labeling; (2) the forced use of a reduced set of synergies might have made the differences more salient. 



In summary, our framework uses feature engineering to estimate AUs as muscle movement sources from distal EMG, overcoming the source muscle identification limitation. It has an advantage over relying only on visual cues because muscle activation does not translate 1:1 to movement kinematics. We used the results to analyze facial muscle synergies (NNMF) in an automatic manner; compared it to CV-only annotated data; and Human-annotated data, to understand the potential and limitations of this method. We conducted the co-occurrence analysis using coder’s data, CV, and EMG. The objective was to find common factors in how a smile is produced, and the relationship between visible movement and EMG synergies.

\section{Limitations of this study}
One of the limitations of this study was the number of electrodes provided in the EMG wearable. Whilst four electrodes provide a good trade-off between wearability, smile and AU detection, they are limited to conduct muscle synergy analysis. Therefore, we opted to determine the optimal number of synergies using CV only, and a synergy-matching strategy between EMG and CV. Increasing the electrode number will enable us to explore synergies containing more facial expressions. In this case, we opted to model mainly AU6 and AU12, and to consider other AUs in the EMG as ``noise''. Furthermore, synergy analysis based on NNMF requires continuous labels. The human-labeled AUs were performed frame-by-frame, but the coders only indicated presence, not intensity. Thus, we opted to do a block analysis on single and simultaneous activation. Labeling AU intensity as well would have been useful to apply our NNMF method directly on the ground-truth labels. Fortunately, OpenFace derives both continuous and binary AU labels which were useful for our method. However, the CV-only detection still could be improved. In particular, we found out that CV might have confused AU10 for AU12; and AU6 for AU7. Finally, we conducted this study on data aimed to elicit smiles. It would be interesting to assess the synergies present in other types of facial expressions. 

\section{Conclusions and future directions}
Our Sensing-Source-Synergy approach led to good AU identification from a multimodal system, and aided to fine-grained AU synergy analysis. Our results suggest that the Duchenne marker can be displayed in both posed and spontaneous smiles, but the underlying synergy differs. However, these results should be assessed carefully. Our method is computationally inexpensive and performs better estimating AUs than OpenFace alone. The EMG already contains the mixed activity of the movement sources measured distally. ICA estimates them blindly. OpenFace’s AU dynamics are inputs for cross-correlation with EMG, leading to a CV-aided AU label approximation. The final outcome is determined by a threshold on the EMG activity of the approximated source. Accordingly, the EMG AU detection performance is not constrained by OpenFace’s detection accuracy. Nevertheless, the agreement with human coders (i.e., the ground truth) can still be improved. Future work should aim to achieve better agreement to refine the synergy analysis presented here. Another alternative would be to use our proposed algorithm only with human-coded labels, instead of the OpenFace algorithm, in a cross-validation schema that ensures generalizability of the results, given that the labels would be also the ground truth. 


%





\ifCLASSOPTIONcaptionsoff
  \newpage
\fi



\bibliographystyle{IEEEtran}
\bibliography{references}
%



%

\begin{IEEEbiography}[{\includegraphics[width=1in,height=1.25in,clip,keepaspectratio]{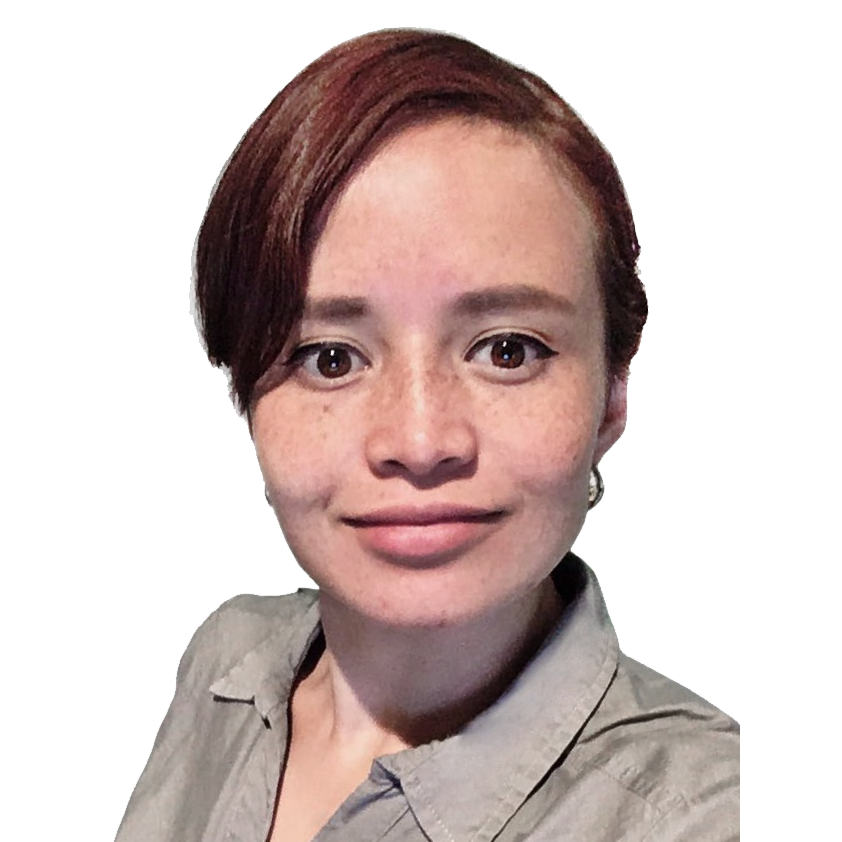}}]{Monica~Perusqu\'ia-Hern\'andez} 
received the BSc in electronic systems engineering (2009) from the Tecnologico de Monterrey, Mexico; the MSc in human-technology interaction (2012) and the professional doctorate in engineering in user-system interaction (2014) from the Eindhoven University of Technology, the Netherlands. In 2018, she obtained the PhD in Human Informatics from the University of Tsukuba, Japan. Her research interests are affective computing, biosignal processing, augmented human technology, and artificial intelligence.
\end{IEEEbiography}

\begin{IEEEbiography}[{\includegraphics[width=1in,height=1.25in,clip,keepaspectratio]{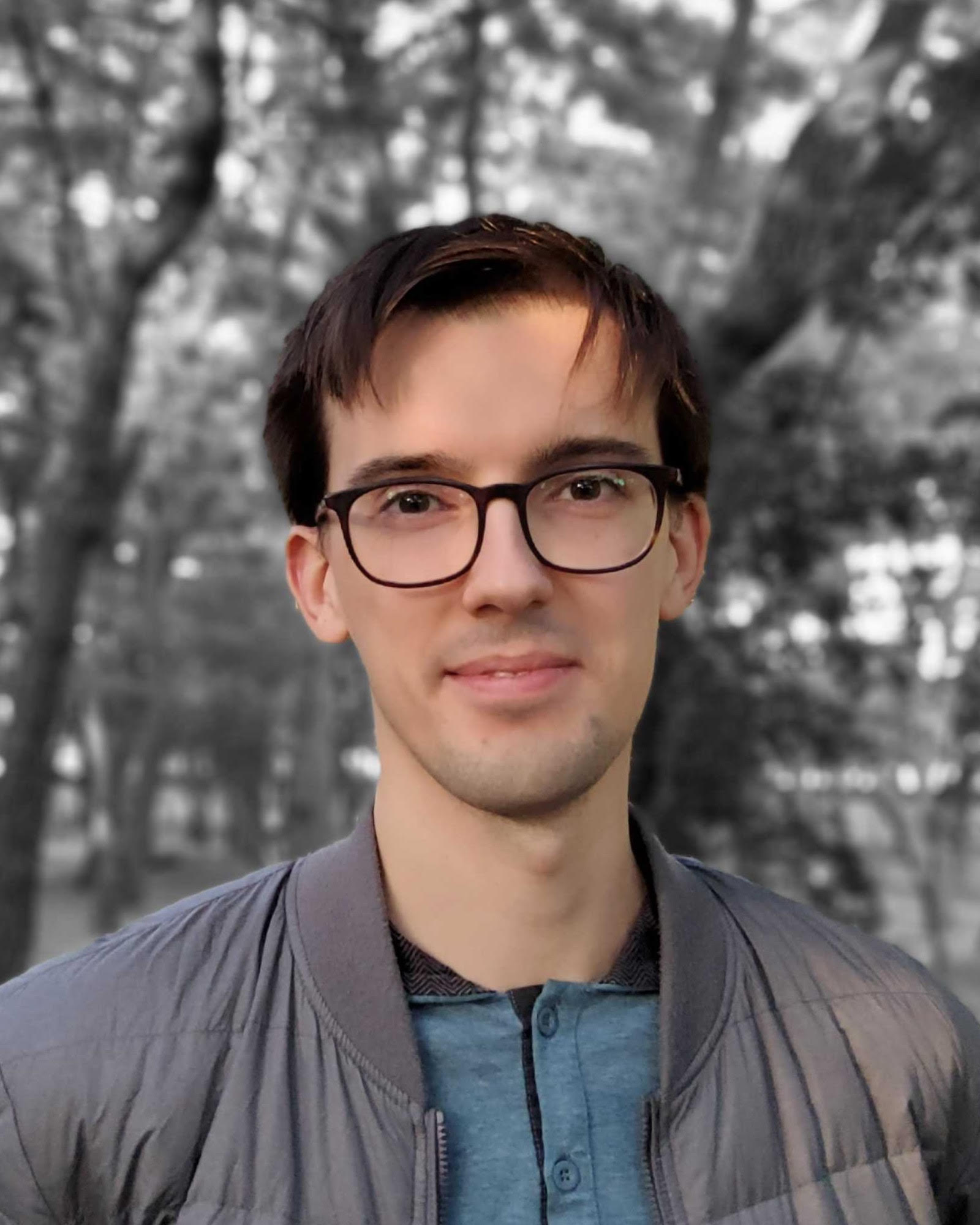}}]{Felix Dollack}
received the BEng degree in hearing technology and audiology from the Jade University of Applied Science in Oldenburg, Germany, in 2013 and the MSc degree in hearing technology and audiology from the Carl-von-Ossietzky University in Oldenburg, Germany, in 2015. In 2020, he obtained the PhD degree in Human Informatics at the Empowerment Informatics Program, University of Tsukuba, Japan. His research interests include psychophysics, biosignal processing and artificial intelligence.
\end{IEEEbiography}


\begin{IEEEbiography}[{\includegraphics[width=1in,height=1.25in,clip,keepaspectratio]{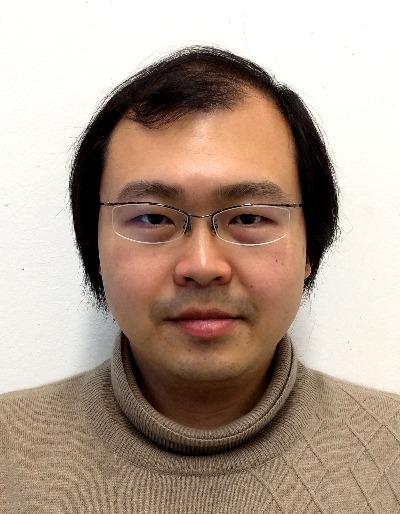}}]{Chun Kwang Tan}
received his MSc in Autonomous Systems from the University of Applied Science Bonn-Rhein-Sieg, Sank Augustin, Germany, in 2016. In 2020, he received PhD degree in Human Informatics at the Emowerment Informatics Program, University of Tsukuba, Japan. His research interests includes EMG analysis, gait analysis and neural motor control.
\end{IEEEbiography}

\begin{IEEEbiography}[{\includegraphics[width=1in,height=1.25in,clip,keepaspectratio]{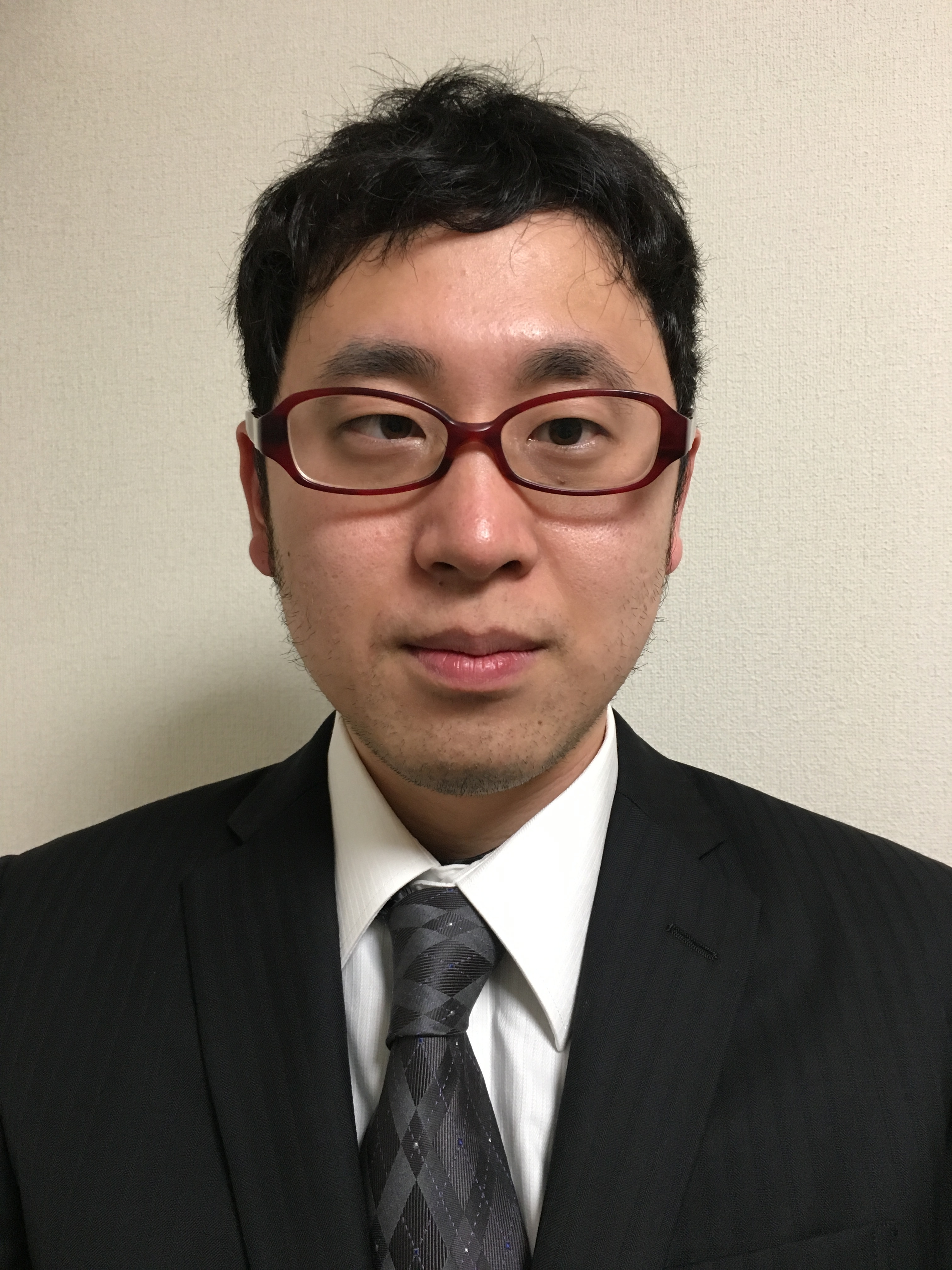}}]{Shushi Namba}
received the Ph.D. degree in psychology from Hiroshima University, Japan, in 2018. He is currently Assistant Professor of Hiroshima University. His research interests include genuine facial displays, human computing and social cognition.
\end{IEEEbiography}

\begin{IEEEbiographynophoto}{Saho Ayabe-Kanamura}
received the PhD degree in psychology from the University of Tsukuba, Japan, in 1998. Since 2013, she is professor of the Faculty of Human Sciences at the University of Tsukuba. Her research interest includes human chemosensory perceptions, perceptual learning and cognitive neurosciences.
\end{IEEEbiographynophoto}

\begin{IEEEbiography}[{\includegraphics[width=1in,height=1.25in,clip,keepaspectratio]{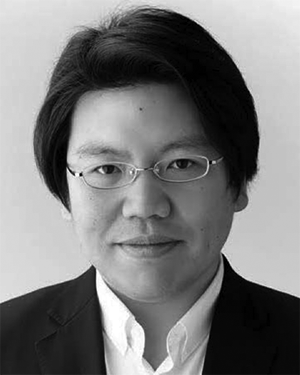}}]{Kenji~Suzuki} received the Ph.D. degree in pure and applied physics from Waseda University, Tokyo, Japan, in 2003. He is currently Professor with the Faculty of Engineering, Information and Systems, and also Principal Investigator with the Artificial Intelligence Laboratory, University of Tsukuba, Tsukuba, Japan. Prior to joining the University of Tsukuba in 2005, he was a Research Associate with the Department of Applied Physics, Waseda University. He was also a Visiting Researcher with the Laboratory of Physiology of Perception and Action, College de France, Paris, France, in 2009, and with the Laboratory of Musical Information, University of Genoa, Genoa, Italy, in 1997. His research interests include artificial intelligence, cybernetics, assistive and rehabilitation robotics, computational behavior science, and affective computing.
\end{IEEEbiography}





\end{document}